\documentclass{article}
\usepackage{equation,epsfig}
\begin{document}
\baselineskip 0.55cm
%%%
%\draft
\pagestyle{empty}
\begin{flushright}
{CERN-TH/2000-093}\\
{hep-ph/0003260}
\end{flushright}
\begin{center}
{\Large \bf Single top quark $tW+X$ production at the LHC: a closer look}\\
\vspace*{1cm} 
{\bf     A.        Belyaev    $^{1}$$^{*)}$,  
         E.        Boos       $^{2}$} 
\\
\vspace*{0.3cm} 
          $^1$ CERN Theory Division, CH--1211 Geneva 23, Switzerland\\
          $^2$ Skobeltsyn Institute for Nuclear Physics,
               Moscow State University, 119 899, Moscow, Russia\\
\vspace*{2cm}  
{\bf ABSTRACT} \\ \end{center}
\vspace*{5mm}
\noindent
We have reexamined  the $tW+X$ single 
top-quark production  process which is important
at the LHC contrary to the Tevatron.  
The special attention was paid  to the treatment of 
the $2\to 2$[$Wt$] process
and the part of it's next-to-leading correction:   $2\to 3$[$Wtb$] process.
We show that $2\to 3$[$Wtb$] process has to be correctly taken 
into account with a proper subtraction of the top pair contribution and
that it has qualitatively different  kinematical distributions
from  the $2\to 2$[$Wt$] process. 
We present  the total cross section of the $tW+X$
production to be about 62 pb at QCD scale 
be taken as a top quark mass, suggest the method
of combining $Wt$ and $Wtb$ processes
with gauge invariant subtraction of the $t \bar t$ part
which allows to reproduce correct kinematical properties  and 
perform a proper event simulation of the $tW+X$ process  
in the whole kinematical region.
 
\vspace*{5cm} 
\noindent 
\rule[.1in]{16.5cm}{.002in}

\noindent
$^{*)}$   On leave of absence from     Skobeltsyn Institute for Nuclear Physics,
          Moscow State University, 119 899, Moscow, Russia 
\vspace*{0.5cm}
\begin{flushleft} 
CERN-TH/2000-093 \\
March 2000
\end{flushleft}
\vfill\eject
%\pagestyle{empty}
%\clearpage\mbox{}\clearpage

\newpage
\setcounter{page}{1}
\pagestyle{plain}

%%%%%%%%%%%%%%%%%%%%%%%%%%%%%%%%%%%%%%%%%%%%%%%%%%%%%%%%%%%%%%%%%%%%%%%%%

\section{Introduction}
The study of the electroweak single top-quark physics
is a very important  part of research programs at future TeV energy
colliders.
Such a study allows to 
investigate with high enough accuracy properties of the top-quark and
to measure a $Wtb$ coupling structure. It may shed a light on the
underlying theory which probably stands beyond the  Standard Model
~\cite{previous,boos1,young,Boos:1999dd}. Besides, single top
production at LHC has a large rate of the order of 300 pb and
therefore it gives an important part of the background to 
various "new physics" processes. 

In this paper we concentrated on the $pp\to tW+X$ production
process at the LHC. This process was the subject of the previous 
studies~\cite{Moretti:1997ng,Belyaev:1999dn,Tait:2000cf}. 
In the paper~\cite{Belyaev:1999dn} we  have
calculated $tW+X$
process among the others processes important at the Tevatron and LHC colliders.
In order to separate  $tWb$ process of the single top-quark production from the
$t \bar t$ pair  production we have introduced a cut on the invariant $Wb$
mass window $('approach \ I')$. In the paper ~\cite{Tait:2000cf}
the cross section of $tW$ process was calculated  in
a different approach where  $t \bar{t}$ contribution was subtracted from
$tWb$ process  in the narrow width approximation ($'approach \ II'$).
However, both approaches have different drawbacks and  have
some aspects which have been treated not quite correctly.
Therefore it deserves 
closer look at the process.

In $'approach \ I'$ results are formally  $|m_{WB}-m_{top}|$ cut
dependent. In the paper~\cite{Belyaev:1999dn} this cut was chosen too
modest from the experimental point of view ($|m_{WB}-m_{top}|<3
\Gamma_t$) since the real experimental  mass window for the top-quark is
of the order of 20 GeV $\simeq 10-15 \Gamma_t$. This was pointed out in
the paper ~\cite{Tait:2000cf}. However, in the present  study we give 
the pure theoretical arguments how the cut should be 
chosen and explain why it has to be significantly larger
than the top width. This cut should be of the order of $~20$~GeV
 even in case of an ideal
detector. After that it will be clear that a formal cut dependence
is significantly reduced. Specially one should stress that the 
($'approach \ I'$) reproduces $correctly$ kinematical distributions
   which is the crucial point for a phenomenological analysis.

The  $'approach \ II'$ in it's turn is cut independent but
     it does not have receipt how to simulate $Wtb$ events at all.
    In the paper ~\cite{Tait:2000cf} 
the only $2\to 2$[$Wt$] process
($b g \rightarrow W t$) has been used for an event simulation. 
However in this case
an important part of $Wtb$ events are absent
and such an implementation of the $'approach \ II'$ leads to
the wrong kinematical distributions for  $tW+X$ process.  
For the  numerical values 
of the cross section the QCD scale $\hat{s}$ has been used.
But one should point out that 
 such a scale is too large and leads to significantly lower 
rate even for top quark pair production as we know from NLO
calculations \cite{Bonciani:1998vc}. For single top $Wt$ process 
NLO results have not been obtained yet, however one would expect
lower characteristic scale comparing to top pair.

In this paper we have been developed the $'approach I'$.
We apply the reasonable $|m_{WB}-m_{top}|$ cut
consistent with theoretical arguments and an experimental mass
resolution for top-quark.
We have calculated also the cross sections using 
$'approach II'$ for the cross check.

For both methods one needs to apply subtraction procedure to avoid  double
counting.  We have
suggested the new  procedure of  the  combining of the two different processes. This
method  implies the correct subtraction
procedure,  reproduces  the correct kinematical properties in the whole
phase space region, and  gives stable results.

Our paper is organised as follows.
In the section II  we compare two different approaches
of a subtraction of $t\bar t$ pair production from 
the process with $tWb$ final state.
In section III we develop the new approach for
a treatment of double counting and combining two signal processes together.
 In section IV we present the final results
and draw the conclusions.
%%%%%%%%%%%%%%%%%%%%%%%%%%%%%%%%%%%%

\section{Leading order $2\to 2$ and 
{$\cal O$}($1/log m_t^2/m_b^2$) $2\to 3$ process:
subtraction of $t\bar{t}$ pair production}

\begin{figure}[htb]
\noindent
\mbox{\epsfig{file=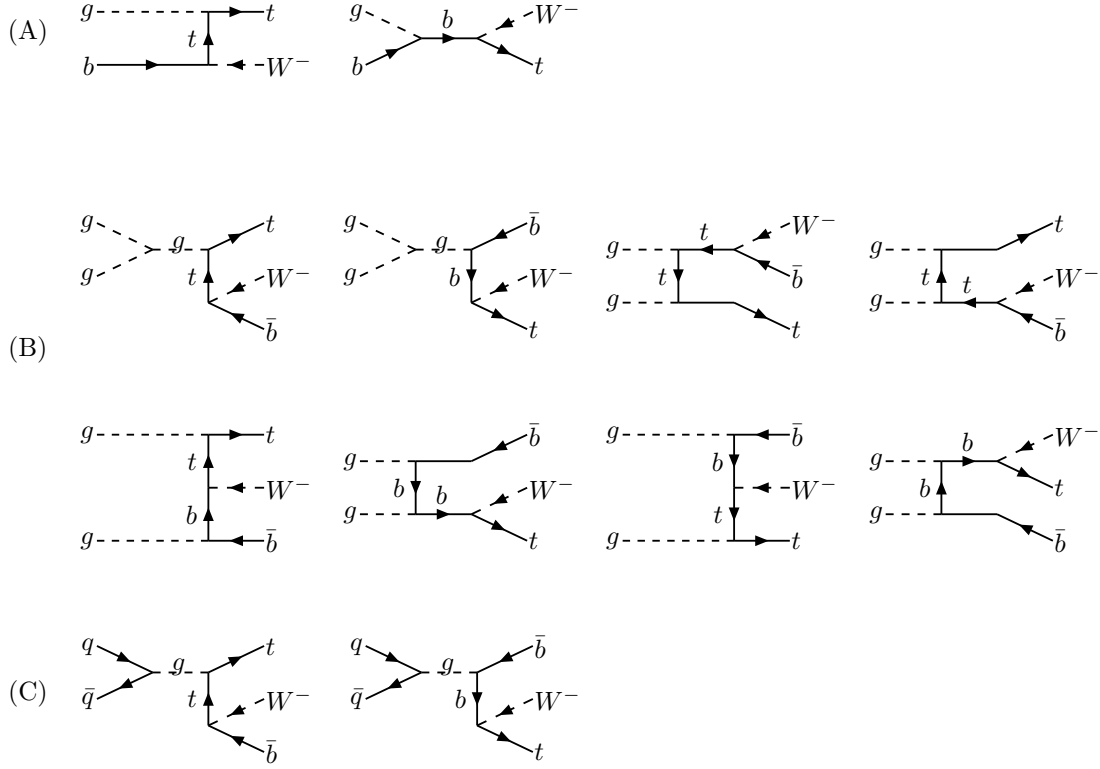}}
\caption{\label{diag_tw} Diagrams for leading order $2\to 2$ 
$tW$ production  (A) and {$\cal O$}($1/log m_t^2/m_b^2$) $2\to 3$ process
(B), (C)}
\end{figure}

In Fig.~\ref{diag_tw} we present the complete gauge invariant set of leading
order and {$\cal O$}($1/log m_t^2/m_b^2$) diagrams contributing to the
$tW+X$ final state.

There are two problems one should avoid in order to combine
correctly different contributions to the $tW$ single top production:
one should remove the contribution
from $t\bar{t}$ production, giving the same $tWb$ final state and 
take care about the double counting 
which takes place when  one simply adds contribution from two 
$2\to 2$ and $2\to 3$ processes.

This happens because $2\to 3$ process is singular in the region of collinear
b-quarks coming from gluon splitting. The same singularity
has been resolved for  $2\to 2$ process when b-quark PDF was defined and
collinear contributions of b-quark was resumed.
The  contribution from the collinear region should be taken only once, 
and therefore one
should apply a subtraction procedure. It should be noticed that $2\to 2$
and $2\to 3$  processes have overlapping only for the leading log 
gluon splitting term. 

In this section we would like to 
compare two different approaches of solving the first
problem -- subtracting the contribution from the top-quark pair production.
As for double counting, we use the conventional solution 
in this section~\cite{Belyaev:1999dn,Tait:2000cf}, namely,
we use the  subtraction of gluon splitting term:
\begin{equation}
\sigma(gb + gg \rightarrow tW +X)_{real}=
\sigma(gb\rightarrow tW)+\sigma(gg\rightarrow tW\bar b)-
\sigma(g\rightarrow b\bar b \otimes gb\rightarrow tW)
\end{equation}

As it was mentioned  in the introduction,
 there are two basic approaches to remove the
contribution from top-quark pair production in a gauge invariant way. 
The first ($'approach I'$),  is the application of the cut
on the invariant mass of $Wb$-- pair in order to remove the resonant $t\bar{t}$
contribution. This procedure is cut dependent,  but it has the straightforward
receipt how to simulate single-top quark production events with the proper
kinematics. One should note however that the cut dependence is not arbitrary
since the cut should be applied according to well known mass 
resolution~\cite{:1999fr} which is typically 10-15 GeV. In terms of the 
top-quark width the window  cut should be applied to remove the $t\bar{t}$
contribution would be of the order of $\simeq\pm 20$ GeV $\simeq$ $10-15\Gamma_{top}$.

The second ($'approach II'$) way of subtraction of $t\bar{t}$
contribution is    the narrow width limit approach
~\cite{Tait:2000cf}:
\begin{equation}
\sigma(gg \rightarrow tWb)_{single top}=
\sigma(gg \rightarrow tWb)_{total}- \sigma(gg \rightarrow t\bar{t})*Br(t\to Wb) 
-interf[t\bar{t}\otimes tWb],
\end{equation}
where $interf[t\bar{t}\otimes tWb]$ means interference 
of $t\bar{t}$ diagrams with the non-resonant ones.
This procedure formally should reproduce the correct production rate for the
single top quark. But from the practical point of view, it does not give any
receipt how to simulate events of the single top-quark production which is crucial
for the further
kinematical studies.

Table~\ref{table:cs1} shows the results for two methods of subtraction of the
$t\bar{t}$ contribution. In order to give the idea how strong  is the 
dependence on the  $Wb$  invariant mass we  present numbers  for two -- 
$10$ and $15 \Gamma_{top}$ window cuts  
which corresponds to $\pm 16$ and $24$ GeV mass windows respectively.
All the numerical results have been obtained by means of the
program CompHEP \cite{comphep}.

\begin{figure}[htb]
\noindent
\mbox{\epsfig{file=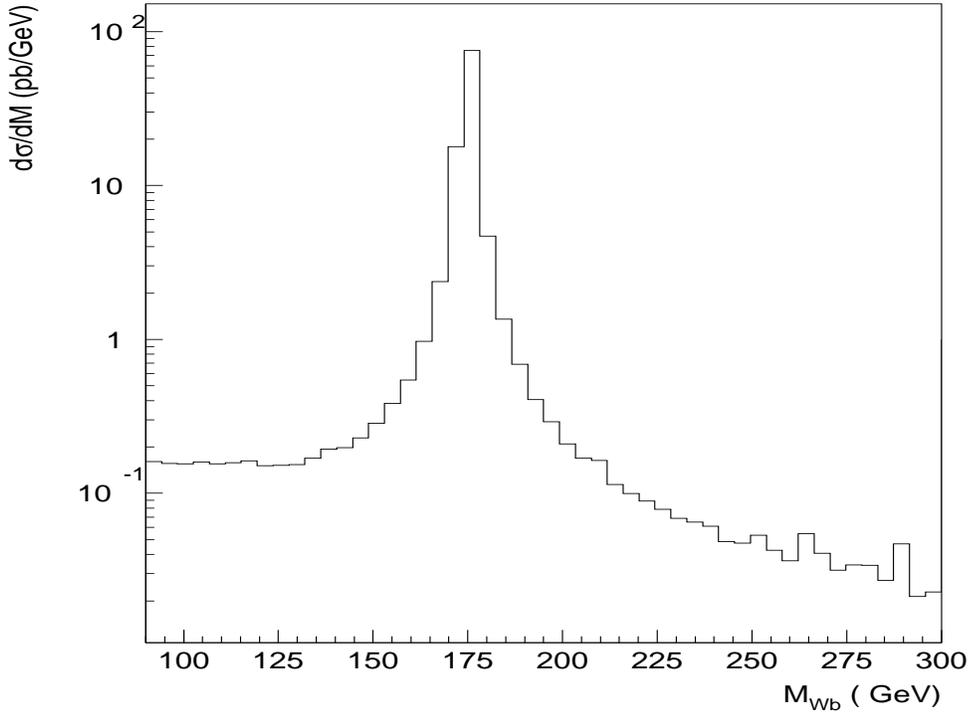 ,width=0.8\textwidth,height=0.6\textwidth}}
\caption{\label{fig:wb} Parton $Wb$ invariant mass distribution for 
$gg\to tW^-\bar{b}$ processes. }
\end{figure}

Results in the table are shown for several characteristic QCD
factorisation/normalisation scales:
$Q=m_{W}, m_{top}, m_{top}+m_{W}$ GeV which
give a natural scale interval for the process under study. We also show 
in the last column the results at $Q^2 = \hat{s}$ for a comparison to 
previous calculations. 
From the table one can see that the subtraction term $g\to b\bar b 
\otimes (gb+bg)\to tW^-$
is of the order of $80\%$ of $(gb+bg)\to W^-t$ cross section.
One can also see that $'approach \ II'$ of subtraction 
of $t\bar{t}$ contribution
gives   cross section
for  $W^-t+X$ process consistent with  the $'approach \ I'$
for $\pm 15 \Gamma_{top}$  $W^-b$ mass window cut.
For example, one  has 31.0 and 28.9 pb respectively for these two methods
at $\mu=m_{top}$.
Since the $'approach \ I'$ is more physical in the sense that it allows 
to reproduce the correct kinematical distribution we use
it with $\pm 15 \Gamma_{top}$  $W^-b$ mass window cut for the final
results.  One could also take a look at the $Wb$ invariant mass distribution 
at the parton level which is presented in Fig.~\ref{fig:wb}.
From this figure one can clearly see that $\simeq 25~GeV \ Wb$ mass window
would completely  remove $t\bar{t}$ contribution with its interference to the $gg\to
tW^-\bar{b}$ processes.

One could easily perform the fitting procedure 
which leads to the more quantitative answer about the size of this window
and gives those 25 GeV. This procedure significantly reduce the ambiguity of the choice
of this window cut which is also 
of the order of the experimental mass resolution mentioned above.

Contribution from $qq\to tWb$ process
has been also taken into account in our study. 
The contribution from this process is not 
negligible and is of the order of 7\% to the $tWb$ final state after 
the removing $t\bar{t}$
contribution.

One should  also notice that cross sections are quite scale dependent. For
three QCD scales: $Q=m_{W}, m_{top}, m_{top}+m_{W}$ GeV 
the uncertainty due to the choice of different scales are
of the order of 25-30\%. The choice of the scale $Q=\sqrt{\hat{s}}$ gives 
significantly lower results. But we should stress that high QCD scale
$Q=\sqrt{\hat{s}}$ seems to be unphysical since it gives almost factor two
lower cross section even for $t\bar{t}$ production at tree level 
in comparison with the next-to-leading(NLO)  order result
~\cite{Bonciani:1998vc}. For  $Q=m_{top}=175$ GeV tree level  result
is much close to NLO one.   Therefore it is
quite reasonable to use  $Q=m_{top}=175$ GeV choice  for the processes
involving single top-quark  production for which the physical scale could be
even smaller then for the  $t\bar{t}$ pair production.

\begin{table}[h]
\begin{tabular}{| l | c | c | c | c |}
PROCESS  			
& \multicolumn{4}{c|}{CS(pb), CTEQ4L}\\
\cline{2-5}
& $\mu=m_{W}$ & $\mu=m_{top}$& $\mu=m_{top}+m_W$ & $\mu=\hat{s}$\\
\hline
$(gb+bg)\to W^-t$ 	  	&29.4	& 29.1	& 28.6	     & 27.8\\
$g\to b\bar b 
\otimes (gb+bg)\to tW^-$  	&25.3	& 23.9  & 22.9	     & 21.8\\
\hline
$gg\to t\bar{t}$	   	&717	& 523	& 457	     & 358 \\
$\mbox{interference}
[t\bar{t}\otimes tWb]$	   	&-14.6	&-10.7	& -9.19	     & -7.14 \\
$gg\to tW^-b$ (no cuts)	  	&737	& 536	& 469	     & 368 \\
$gg\to tW^-b$ 
$(\pm 10 \Gamma_{top} 
\ W^-b$ mass cut)	  	&42.7	& 30.9	& 27.0	     & 20.8 \\
$gg\to tW^-b$ 
$(\pm 15 \Gamma_{top} 
\ W^-b$ mass cut)	  	&33.3	& 24.3	& 21.2	     & 16.5 \\
\hline
\hline
$q\bar{q}\to t\bar{t}$	 	& 99.0  & 79.0	& 72.0	     & 61.4 \\
$q\bar{q}\to tW^-b$ (no cuts)	& 98.6  & 78.7	& 71.6	     & 61.0 \\
$q\bar{q}\to tW^-b$ 
$(\pm 10 \Gamma_{top} 
W^-b$ mass cut)			& 2.9  & 2.3	& 2.1	     & 1.7 \\
$q\bar{q}\to tW^-b$ 
$(\pm 15 \Gamma_{top} 
\ W^-b$ mass cut)	  	& 1.9  & 1.5    & 1.3        & 1.1 \\
%%%%%%%%%%%%%%%%%%%%%%%%%%%%%%%%%%%%%%%%%%%%%%%%%%%%%%%%%%%%%%%%%%%%%%%%%%%%%
\hline
\hline
$(q\bar{q}+gg)\to t\bar{t}$      & 816  &602 	& 529    & 419  \\
$(q\bar{q}+gg)\to tW^-b$(no cuts)& 836  &615	& 541	 & 429  \\
$(q\bar{q}+gg)\to tW^-b$ 
$(\pm 10 \Gamma_{top} 
\ W^-b$ mass cut) 		& 45.6	&33.2	& 29.1	 & 22.5 \\
$(q\bar{q}+gg)\to tW^-b$ 
$(\pm 15 \Gamma_{top} 
\ W^-b$ mass cut)		& 35.2	&25.8	& 22.5	 & 17.6 \\
\hline
%%%%%%%%%%%%%%%%%%%%%%%%%%%%%%%%%%%%%%%%%%%%%%%%%%%%%%%%%%%%%%%%%%%%%%%%%%%%%
\hline
$W^-t \ +  \ 
[W^-tb]_{I(\pm 10 \Gamma )}
\ - \ [g\to b\bar b 
\otimes (gb+bg)\to tW^-$]	& 49.7	& 38.4   & 34.8   & 28.5 \\
{\bf $W^-t \ +  \ 
[W^-tb]_{I(\pm 15 \Gamma )}
\ - \ [g\to b\bar b 
\otimes (gb+bg)\to tW^-$]}	& {\bf 39.3}	& {\bf 31.0}   & {\bf 28.2 } & {\bf 23.6} \\
$W^-t \ +  \ [W^-tb]_{II} \ - \
[g\to b\bar b 
\otimes (gb+bg)\to tW^-$]	&38.3	& 28.9   & 26.9  & 23.1  \\
\hline
\hline
\end{tabular}
\caption{\label{table:cs1} cross section for various 
processes contributing to $tW- +X$ production with and without 
cut on the invariant $Wb$ mass}
\end{table}

\section{Treatment of the double counting: 
comparison and combining of the  $Wt$+ISR  and complete $tWb$ processes}

In this section we would like make close look 
at the solution of the double counting problem.

\begin{figure}[htb]
\noindent
\mbox{\epsfig{file=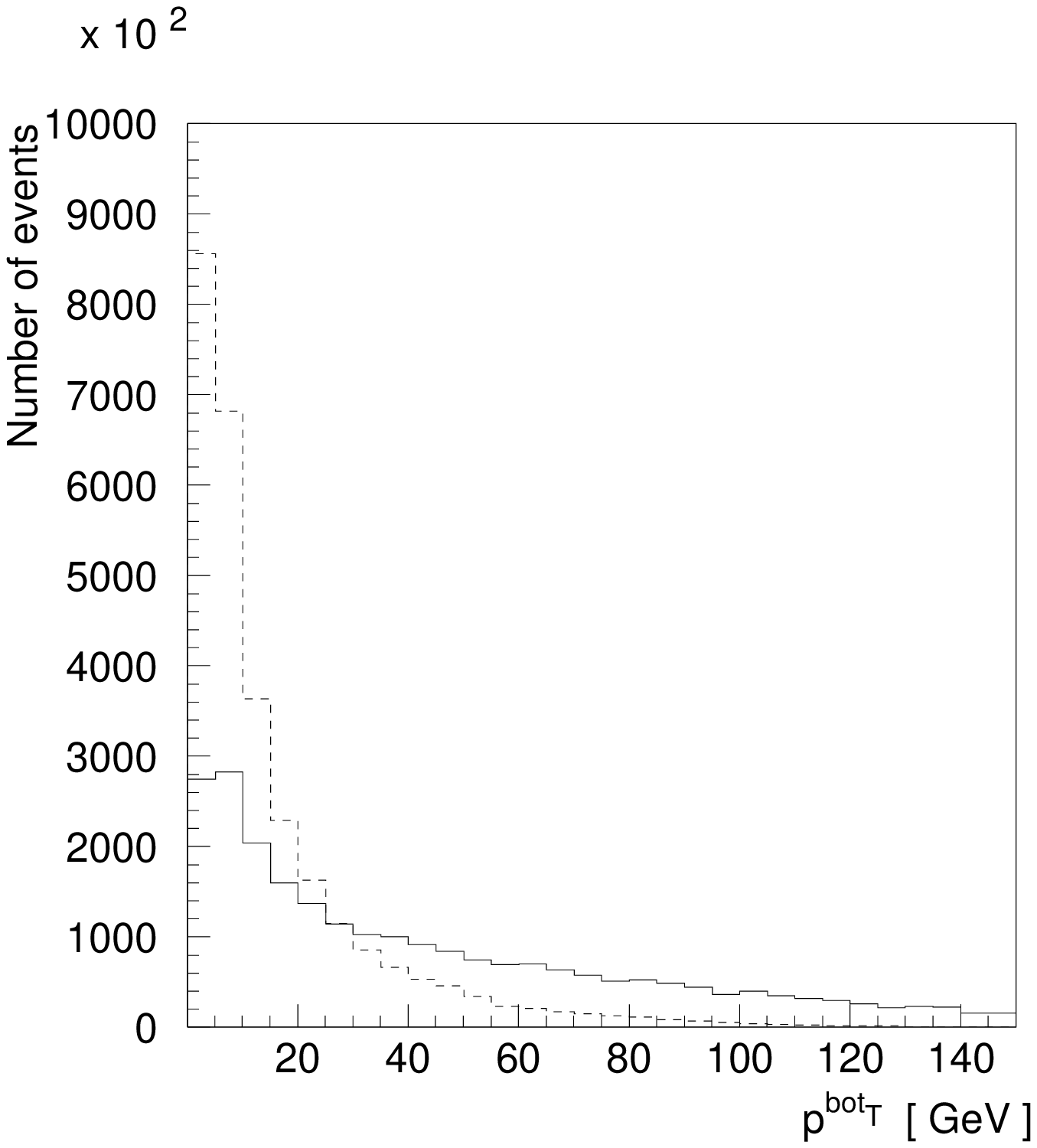 ,width=0.45\textwidth,height=0.35\textwidth}
      \epsfig{file=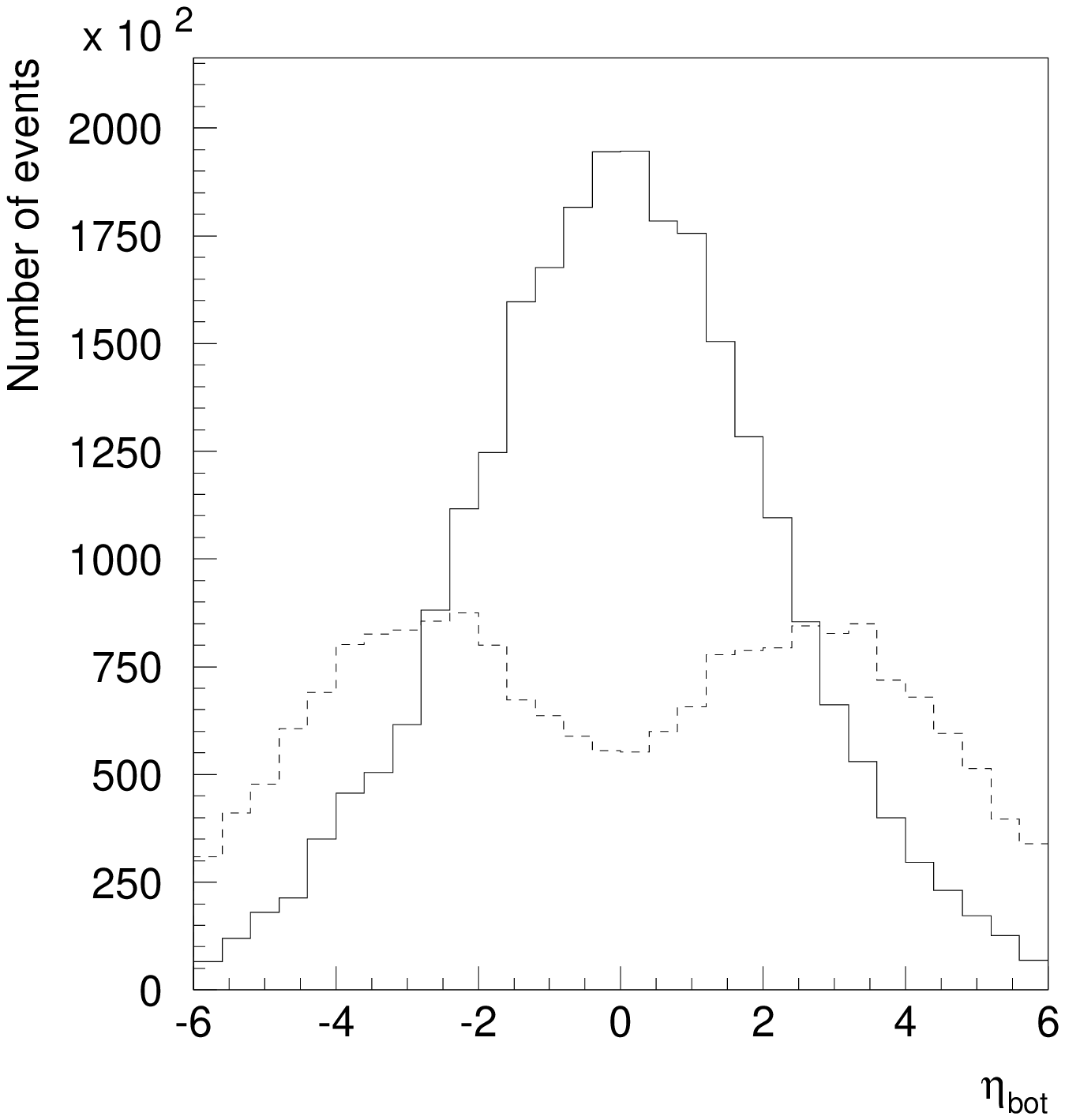,width=0.45\textwidth,height=0.35\textwidth}}
\mbox{\epsfig{file=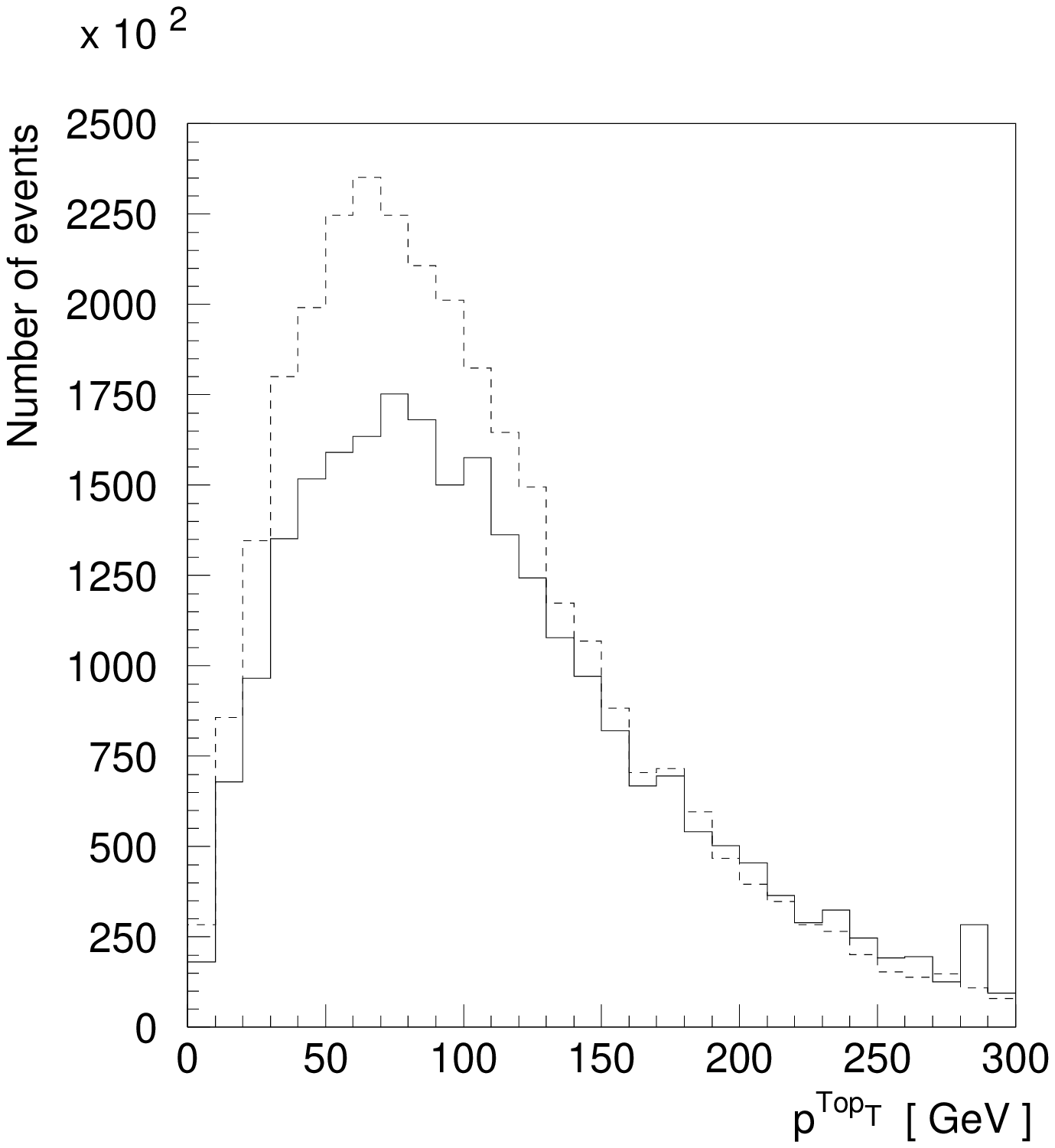 ,width=0.45\textwidth,height=0.35\textwidth}
      \epsfig{file=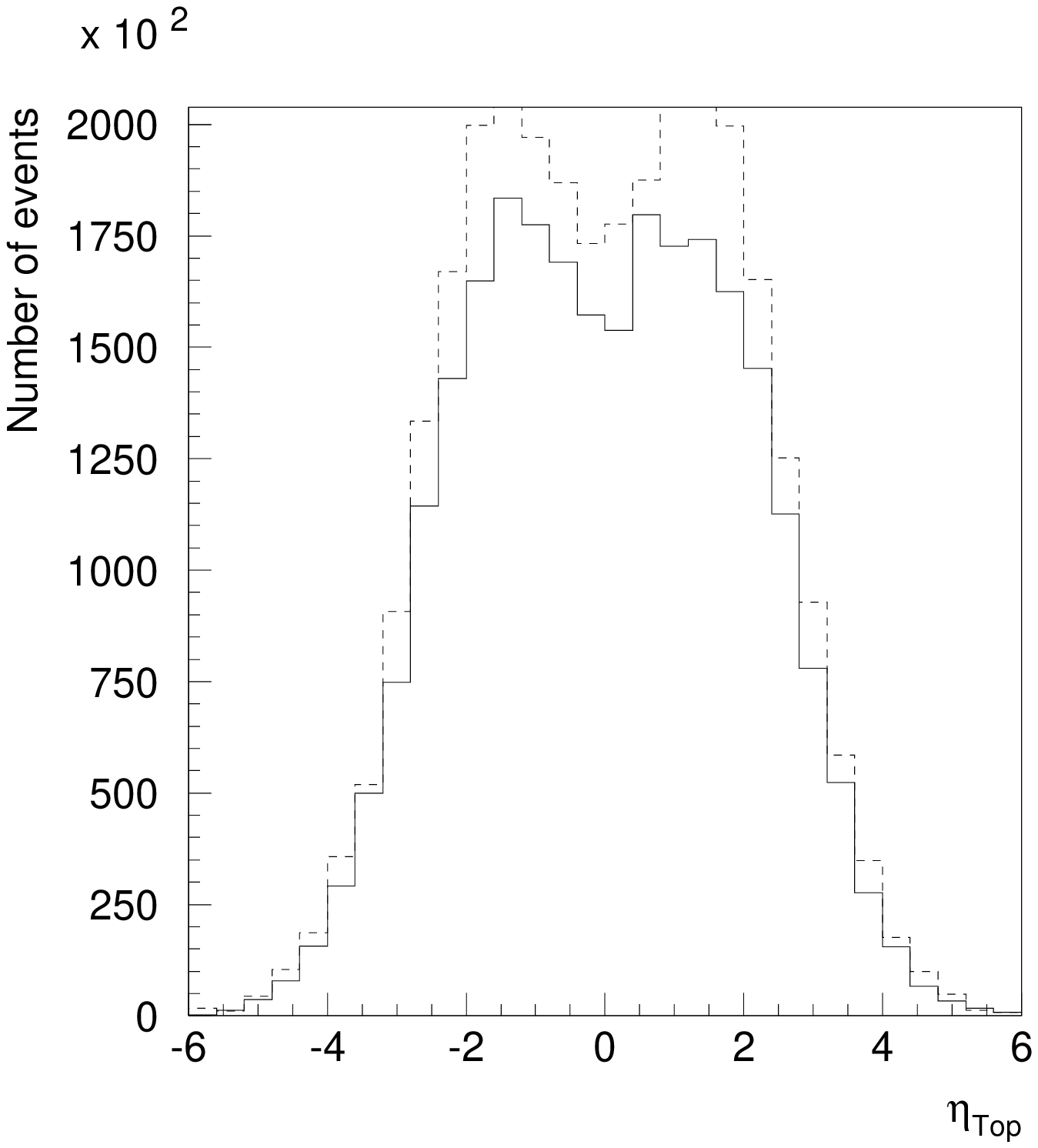,width=0.45\textwidth,height=0.35\textwidth}}
\mbox{\epsfig{file=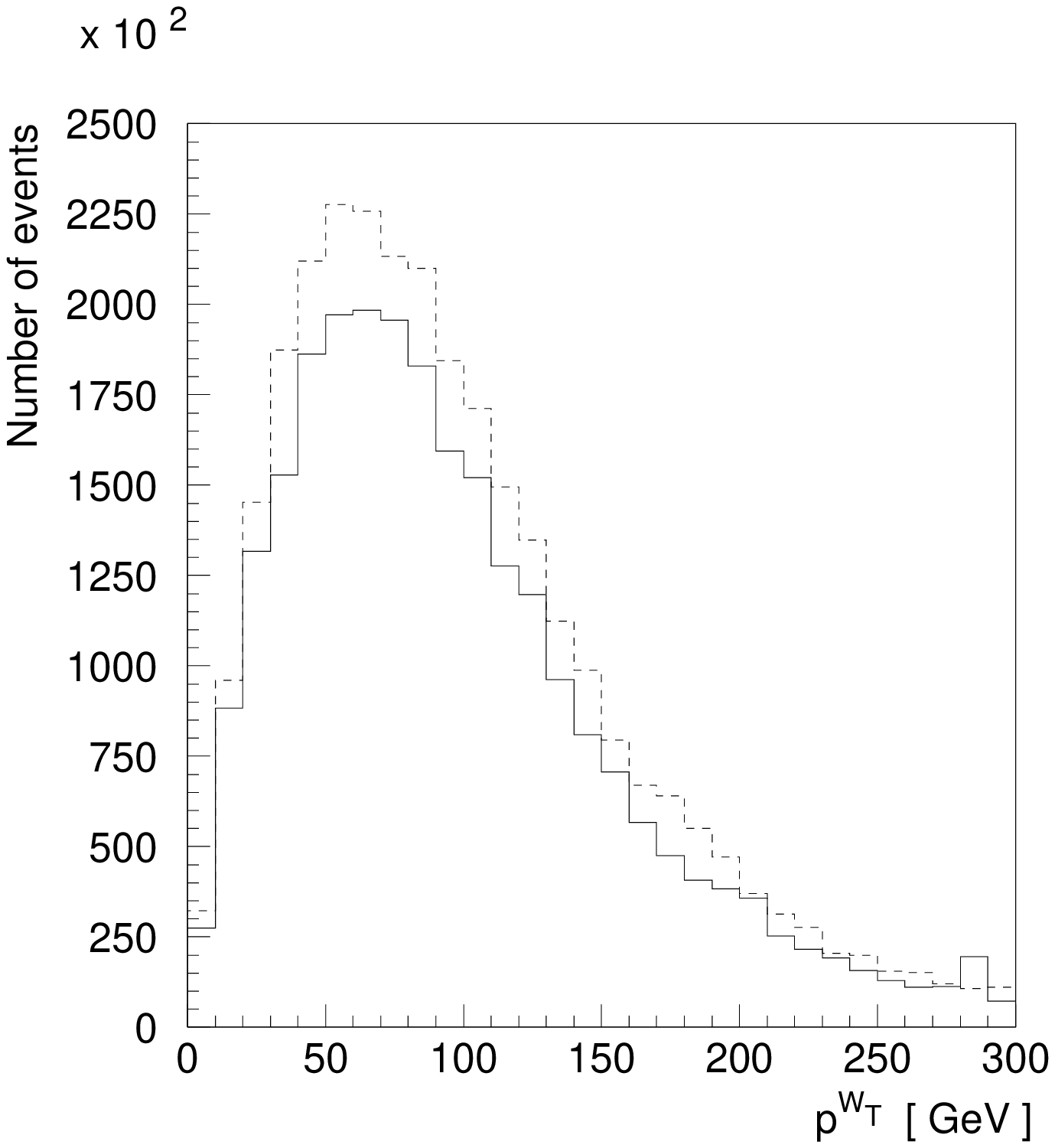 ,width=0.45\textwidth,height=0.35\textwidth}
      \epsfig{file=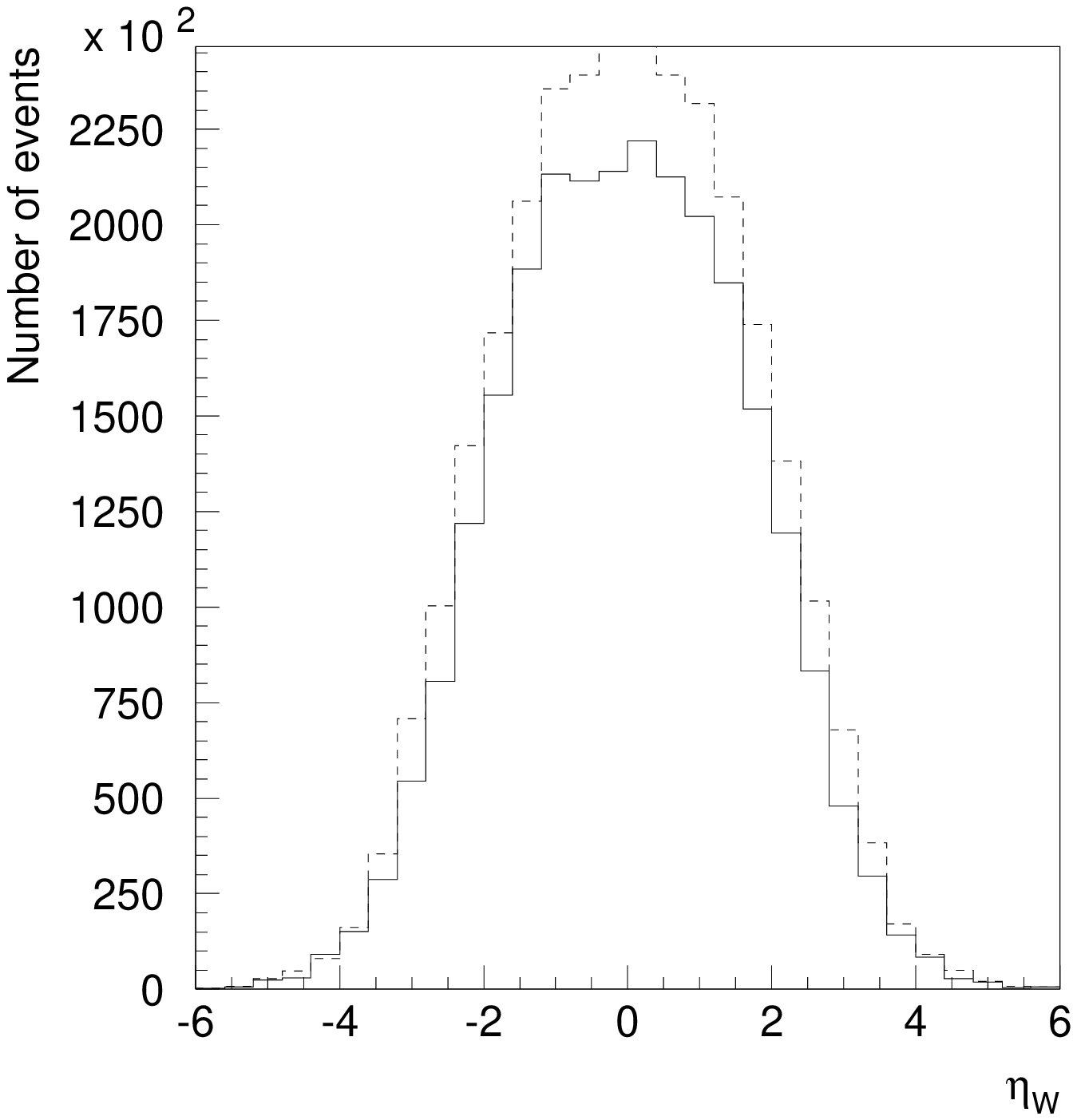,width=0.45\textwidth,height=0.35\textwidth}}
\caption{\label{figure:wt} Transverse momenta and rapidity distributions for
the final state particles of the processes $pp(bg)\to tW-$+ $b_{ISR}$ (dashed line) -- 
$2\to 2$ process with additional b-quark from initial state 
radiation  and true $2\to 3$  $pp(gg+q\bar{q})\to tW^-\bar{b}$ process (solid line). }
\end{figure}

Results from in Table~\ref{table:cs1} have been obtained 
using 'conventional' subtraction procedure.
However one need to study how $2\to 2 $ and $2\to 3$
processes should be combined 
in order to reproduce not only the total cross section
but also the correct event kinematics.
 
We have used the following procedure to work out the receipt for this.
We have compared various kinematical distributions of
$pp(bg)\to tW^-$+ $b_{ISR}$ ---  
$2\to 2$ process with an additional b-quark from the initial state 
radiation  and complete $2\to 3$  --- $pp(gg+q\bar{q})\to tW^-\bar{b}$ process.
In this way one can try to find a proper matching between resumed
contribution at the collinear region for the b-quark and complete 
tree level contribution at the hard region.
Figure~\ref{figure:wt} shows transverse momenta and rapidity distributions
for all three particles in the final state.
As expected one can see the  difference in the b-quark distributions.
For $pp(bg)\to tW^-$+ $b_{ISR}$ process it is much softer and less central
in comparison to the $pp(gg+q\bar{q})\to tW^-\bar{b}$ process.
In the same time one can see that $W-boson$ and $t-$quark distributions are nearly 
the same.

Since we know the absolute value of the combined cross section
we propose the following method to match collinear and hard kinematical regions.
One can  use kinematical $p_T^b$ separation
of $pp(bg)\to tW^-$+ $b_{ISR}$  and $pp(gg+q\bar{q})\to tW^-\bar{b}$ in the regions
$p_T^b<P_T^{cut}$ and $p_T^b>P_T^{cut}$ respectively.

\begin{figure}[tb]
\noindent
\mbox{\epsfig{file=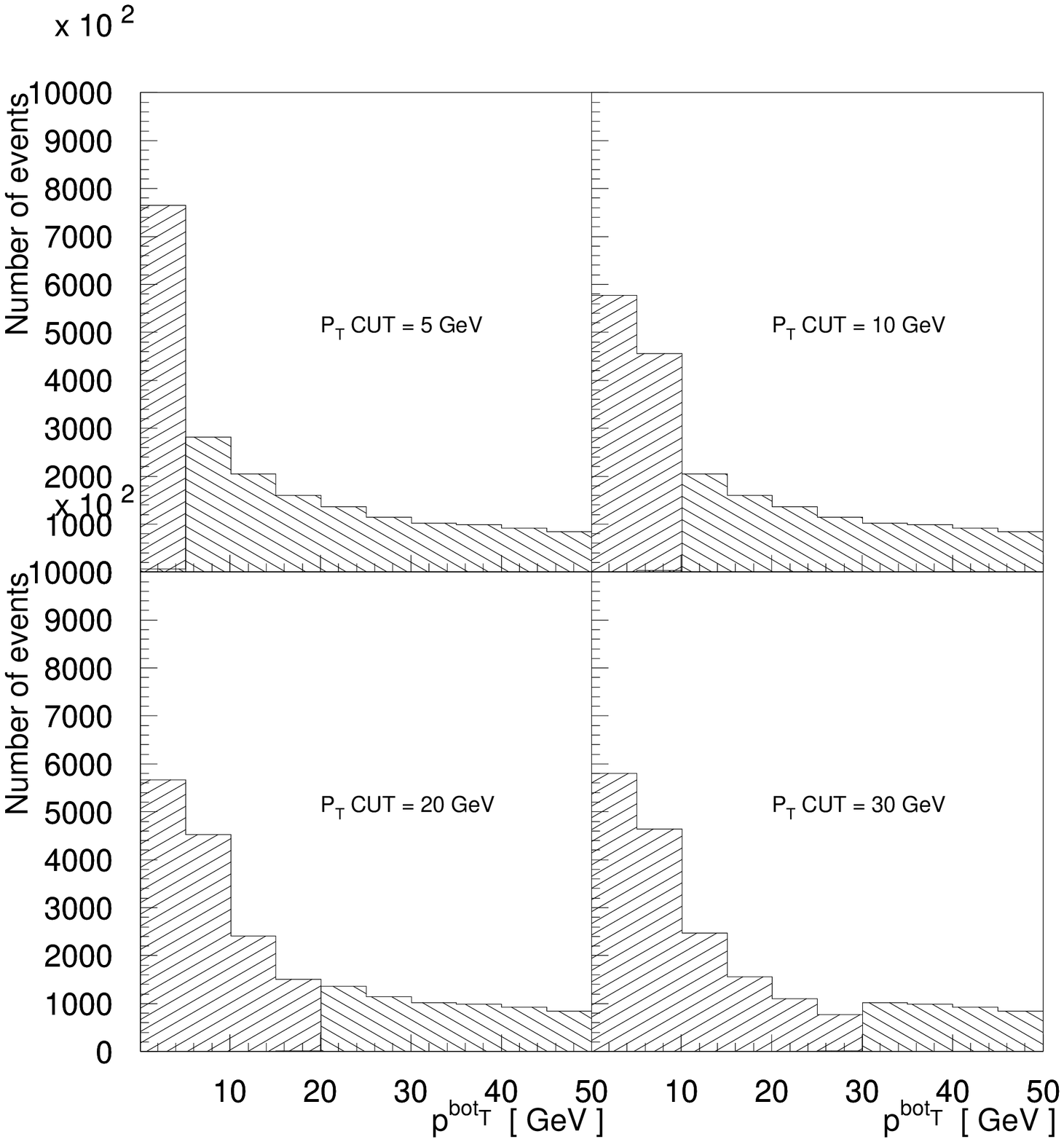 ,width=0.8\textwidth,height=0.6\textwidth}}
\caption{\label{fig:sew} Transverse momenta distribution
of the b-quark sewed for various values of $P_T^{cut}$ for 
 the $pp(bg)\to tW^-$+ $b_{ISR}$ and  $pp(gg+q\bar{q})\to tW^-\bar{b}$ processes. }
\end{figure}

Now one can move the cut and try to satisfy two requirements, namely:\\
1)
the common rate of $pp(bg)\to tW^-$+ $b_{ISR}$ with $p_T^b<P_T^{cut}$
and $pp(gg+q\bar q)\to tW^-\bar{b}$ with $p_T^b>P_T^{cut}$ gives the
combined total rate computed in previous section, 
in other words one can normalise a rate in a collinear region on the
$\sigma_{total} - \sigma[pp(gg+q\bar{q})\to tW^-\bar{b},\ p_T^b>P_T^{cut}]$;\\
2) the overall
$p_T^b$ distribution should be smooth.
\\ 
The result is illustrated in Fig.~\ref{fig:sew} where we show several variants
of combining those two processes for various values of  $P_T^{cut}$.
We have found that the optimal $P_T^{cut}$ providing the smooth sewing
for these two processes at the LHC 
is equal to 20~GeV.  This value  gives physically  
reasonable answer in which regions 
$pp(bg)\to tW^-+ b_{ISR}$  and $pp(gg+q\bar q)\to tW^-\bar{b}$ 
processes should be considered.

We conclude that the method  of combining of the 
$p^b_T$ distribution of 
$Wt$+ISR gluon and complete tree level  $tWb$ process allows to 
find the physically motivated $p_T$ cut on the $b-quark$
which allows us to treat together those processes and 
simulate them in  different kinematical regions of $p^b_T$.

We have estimated  uncertainties due to 
a the choice of the QCD scale within the range $M_W<\mu<M_{TOP}+M_W$
taking the central value of $\mu=M_{TOP}$.
The total cross section presented in Table~\ref{table:cs1} 
is $31.0^{+8.3}_{-1.8}$~pb  within the QCD scale mentioned above.

\section{Final results and conclusions}

We have reexamined  the $tW+X$ single 
top-quark production  process which is important
at the LHC.
We have shown that $2\to 3$[$Wtb$] process has to be correctly taken 
into account with a proper subtraction of the top pair contribution and
that it has qualitatively different  kinematical distributions
from  the $2\to 2$[$Wt$] process. 

We suggest the new method of 'kinematical'
sewing of two different processes contributing to the
$tW+X$ productions using
 the transverse b-quark momenta distribution.
This method allows unambiguously simulate correct kinematical
distribution of the total process of $tW+X$ production
in the whole kinematical region.

We have estimated  the cross section of the
single top  $tW +X$ production taking into account
uncertainties due to the choice of the QCD scale.
The cross section for single top and single anti-top
quark production -- $tW^-+X$  and $\bar{t}W^++X$ at the 
LHC are equal to each other in contrary to other
processes of the single top-quark production.
So, combined [$tW^-+X$ + $\bar{t}W^++X$] cross section 
is $62.0^{+16.6}_{-3.6}$~pb.

\section*{Acknowledgements}
We thank C.-P. Yuan for useful remarks and discussions. 
E.B. is grateful to the Russian Ministry of Science and Technologies,
Russian Foundation for Basic Research (grant 99-02-04011),
INTAS-CERN Foundation (grant No 377), 
and the German Bundesministerium f\"ur Bildung,
Wissenschaft, Forschung und Technologie (BMBF) 
(project no.~HTE0499 Manakos) for  partial financial support.
E.B. would like to thank P. Manakos and Th. Ohl
for their kind hospitality during a visit to Darmstadt University of Technology 
where the paper has been completed.

\end{document}